\documentclass[12pt,a4paper]{article}
\usepackage{amssymb,amsmath}
\usepackage[dvips]{lscape,graphicx}

\newcommand{\bc}{\begin{center}}
\newcommand{\ec}{\end{center}}
\newcommand{\bd}{\begin{displaymath}}
\newcommand{\ed}{\end{displaymath}}
\newcommand{\be}{\begin{equation}}
\newcommand{\ee}{\end{equation}}
\newcommand{\ba}{\begin{array}}
\newcommand{\ea}{\end{array}}
\newcommand{\bt}{\begin{tabular}}
\newcommand{\et}{\end{tabular}}

\newcommand{\ds}{\displaystyle}

\begin{document}


\title{Implementation of the Multiple Point Principle in the
Two-Higgs Doublet Model of type II.}

\author{Colin D.Froggatt${}^{1}$, Larisa Laperashvili${}^{2}$, Roman Nevzorov${}^{3}$,\\
Holger Bech Nielsen${}^{4}$, Marc Sher${}^{5}$\\[5mm]
\itshape{${}^{1}$ Department of Physics and Astronomy,}\\[2mm]
\itshape{Glasgow University, Glasgow, Scotland}\\[2mm]
\itshape{${}^{2}$ Institute of Mathematical Sciences, Chennai, India}\\[2mm]
\itshape{${}^{3}$ School of Physics and Astronomy,} \\[2mm]
\itshape{University of Southampton, Southampton, U.K.} \\[2mm]
\itshape{${}^{4}$ The Niels Bohr Institute, Copenhagen, Denmark}\\[2mm]
\itshape{${}^{5}$ Physics Department, College of William and Mary,} \\[2mm]
\itshape{Williamsburg, USA}
}

\date{}
\maketitle

\begin{abstract}{
\noindent
The multiple point principle (MPP) is applied to the non--supersymmetric two-Higgs doublet
extension of the Standard Model (SM). The existence of a large set of degenerate vacua at
some high energy scale caused by the MPP results in a few relations between Higgs self-coupling
constants which can be examined at future colliders. The numerical analysis reveals that these
MPP conditions constrain the mass of the SM--like Higgs boson to lie below
180 GeV for a wide set of MPP scales $\Lambda$ and $\tan\beta$.
}
\end{abstract}

\vspace{0.5cm}
\footnoterule{\noindent${}^{1}$ c.froggatt@physics.gla.ac.uk\\
${}^{2}$ laper@imsc.res.in \\
${}^{3}$ On leave of absence from the Theory Department, ITEP, Moscow, Russia;\\
$~~$ E-mail: nevzorov@phys.soton.ac.uk\\
${}^{4}$ hbech@alf.nbi.dk\\
${}^{5}$ sher@physics.wm.edu
}

\newpage
\section{Introduction}

The success of the Standard Model (SM) strongly supports the
concept of spontaneous $SU(2)\times U(1)$ symmetry breaking. The
mechanism of electroweak symmetry breaking, in its minimal
version, requires the introduction of a single doublet of scalar
complex Higgs fields and leads to the existence of a neutral
massive particle --- the Higgs boson. Over the past two decades
the upper \cite{1} and lower \cite{1}-\cite{2} theoretical bounds
on its mass have been established. Nevertheless there are good
reasons to believe that the SM with the minimal Higgs content is
not the ultimate theoretical structure responsible for electroweak
symmetry breaking since it is unable to answer many fundamental
questions. For example, if the SM is embedded in a more
fundamental theory characterized by a much larger energy scale
(e.g. the Planck scale $M_{Pl}\approx 10^{19}\,\mbox{GeV}$) than
the electroweak scale, then the Higgs mechanism suffers from a
stability crisis. Indeed, due to the quadratically divergent
radiative corrections, the Higgs boson tends to acquire a mass of
order of the largest energy scale. Low--scale supersymmetry (SUSY)
stabilizes the scale hierarchy, removing quadratic divergences. The
unification of gauge coupling constants, which takes place in
these models at high energies \cite{5}, is commonly considered as
a manifestation of the ultimate underlying theory (e.g.
superstring theory) accommodating gravity. However, the
cosmological constant in SUSY models where supersymmetry is softly
broken diverges quadratically, and enormous fine-tuning is
required to keep its size around the observed value \cite{6}.
Theories with flat \cite{7} and warped \cite{8} extra spatial
dimensions allow one to explain the hierarchy between the
electroweak and Planck scales. They also provide new insights 
into gauge coupling unification \cite{9} and the cosmological 
constant problem \cite{10}.

In this article we exploit the most economical approach 
addressing the hierarchy problem
--- the multiple point
principle (MPP) \cite{11}, which does not require many new
particles or extra dimensions to resolve this problem. MPP
postulates the coexistence in Nature of many phases allowed by a
given theory. It corresponds to a special (multiple) point on the
phase diagram of the considered theory where these phases meet. At
the multiple point the vacuum energy densities of the neighbouring
phases are degenerate.

The multiple point principle applied to the pure SM exhibits a
remarkable agreement with the top quark mass measurements.
According to the MPP, the Higgs effective potential of the SM 
\be
V_{eff}(\phi)=-m^2(\phi)\phi^2+\ds\frac{\lambda(\phi)}{2}\phi^4\,,
\label{1} 
\ee 
which depends only on the norm of the Higgs field
$\phi=\left(\chi^{+},\chi^0\right)$, has two rings of minima in the
Mexican hat with the same vacuum energy density \cite{12}. The
radius of the little ring equals the electroweak vacuum
expectation value (VEV) of the Higgs field. The second vacuum was
assumed to be near the Planck scale $\phi\approx M_{Pl}$.

The mass parameter in the effective potential (\ref{1}) has to be
of the order of electroweak scale ensuring the phenomenologically
acceptable Higgs vacuum expectation value for the physical (first)
vacuum. Since at high scales the $\phi^4$ term in Eq.(\ref{1})
strongly dominates the $\phi^2$ term the derivative of
$V_{eff}(\phi)$ near the Planck scale takes the form: 
\be
\ds\frac{d V_{eff}(\phi)}{d \phi}\Biggl|_{\phi=M_{Pl}}\approx
\left(2\lambda(\phi)+\frac{1}{2}\beta_{\lambda}\right)\phi^3\,,
\label{2} 
\ee 
where $\beta_{\lambda}(\lambda(\phi),g_t(\phi),g_i(\phi))=\ds\frac{d\lambda(\phi)}{d\log\phi}$
is the beta--function of $\lambda(\phi)$, which depends on
$\lambda(\phi)$ itself, gauge $g_i(\phi)$ and top quark Yukawa
$g_t(\phi)$ couplings. Then the degeneracy of the vacua means that
at the second vacuum the Higgs self--coupling and its derivative
must be zero to very high accuracy.

When the Higgs self--couplings tends to zero at the Planck scale,
the corresponding beta--function vanishes only for a unique value
of the top quark Yukawa coupling. Thus by virtue of MPP
$\lambda(M_{Pl})$ and $g_t(M_{Pl})$ are determined. One can then
compute quite precisely the top quark (pole) and Higgs boson
masses using the renormalization group flow (see \cite{12}): 
\be
M_t=173\pm 4\,\mbox{GeV}\, ,\qquad M_H=135\pm 9\, \mbox{GeV}\, .
\label{3} 
\ee 
Shifting the Higgs field VEV in the second vacuum
down from the Planck scale by a few orders of magnitude decreases
the values of the top--quark and Higgs masses, spoiling the
agreement with the experimental data. The hierarchy between the
electroweak and Planck scales might also be explained by MPP
within the pure SM, if there exists a third degenerate vacuum
\cite{13}.

The relationships between different couplings required by MPP
could arise dynamically. For example a mild form of locality
breaking in quantum gravity, due to baby universes say \cite{14},
may precisely fine-tune the couplings so that several phases with
degenerate vacua coexist \cite{book}. However a necessary
ingredient of most models unifying gravity with other gauge
interactions is supersymmetry. At the same time couplings in SUSY
models are adjusted by the supersymmetry so that all global vacua
are degenerate providing another possible origin for the MPP. In
previous papers \cite{15},\cite{151} the MPP assumption has been
adapted to models based on $(N=1)$ local supersymmetry --
supergravity, that allowed an explanation for the small deviation
of the cosmological constant from zero.

As the low--energy limit of an underlying SUSY theory the SM looks
rather artificial. Indeed in order to give masses to all bosons
and fermions in a manner consistent with supersymmetry at least
two Higgs doublets must be introduced. It seems unnatural to
assume that one of them remains light while another acquires a
huge mass of the order of the cut--off scale $\Lambda$
($\Lambda\lesssim M_{Pl}$). Therefore in this article we study the
non-supersymmetric two Higgs doublet extension of the SM
\cite{211},\cite{16} supplemented by the MPP assumption, bearing
in mind supersymmetry as a possible origin of the MPP. In the next
section the SUSY inspired two Higgs doublet model of type II is
outlined and the vacuum stability conditions in this model are
specified. In section 3 the MPP conditions are formulated and the
ensuing relations between the Higgs self--couplings due to the MPP
assumption are established. The Higgs spectrum in the MPP inspired
two Higgs doublet model is discussed in section 4. The
restrictions on the Higgs self--couplings and the SM--like Higgs
boson mass caused by the MPP are explored in section 5. Section 6
contains our conclusions and outlook.

\section{Higgs boson potential and vacuum stability conditions}

The most general renormalizable $SU(2)\times U(1)$ gauge invariant
potential of the the two Higgs doublet model is given by 
\be
\begin{array}{c}
V_{eff}(H_1, H_2)=m_1^2(\Phi)H_1^{\dagger}H_1+m_2^2(\Phi)H_2^{\dagger}H_2-\biggl[m_3^2(\Phi) H_1^{\dagger}H_2+h.c.\biggr]+\\[3mm]
+\ds\frac{\lambda_1(\Phi)}{2}(H_1^{\dagger}H_1)^2+\frac{\lambda_2(\Phi)}{2}(H_2^{\dagger}H_2)^2+
\lambda_3(\Phi)(H_1^{\dagger}H_1)(H_2^{\dagger}H_2)+\lambda_4(\Phi)|H_1^{\dagger}H_2|^2+\\[3mm]
\ds+\biggl[\frac{\lambda_5(\Phi)}{2}(H_1^{\dagger}H_2)^2+\lambda_6(\Phi)(H_1^{\dagger}H_1)(H_1^{\dagger}H_2)+
\lambda_7(\Phi)(H_2^{\dagger}H_2)(H_1^{\dagger}H_2)+h.c. \biggr]\, ,
\end{array}
\label{4}
\ee
where
$$
H_n=\left(
\ba{c}
\chi^+_n\\[2mm]
(H_n^0+iA_n^0)/\sqrt{2}
\ea
\right)\,.
$$
It is easy to see that the number of couplings in the two Higgs
doublet model (2HDM) compared with the SM grows from two to ten.
Furthermore, four of them $m_3^2$, $\lambda_5$, $\lambda_6$ and
$\lambda_7$ can be complex, inducing CP--violation. In what
follows we suppose that mass parameters $m_i^2$ and Higgs
self--couplings $\lambda_i$ of the effective potential (\ref{4})
depend only on the overall sum of the squared norms of the Higgs
doublets, i.e.
$$
\Phi^2=\Phi_1^2+\Phi_2^2\,,\qquad
\Phi_i^2=H_i^{\dagger}H_i=\frac{1}{2}\biggl[(H_i^0)^2+(A_i^0)^2\biggr]+|\chi_i^+|^2\,.
$$
The running of these couplings is described by the 2HDM
renormalization group equations (see \cite{171}--\cite{17}) where
the renormalization scale is replaced by $\Phi$.

At the physical minimum of the scalar potential (\ref{4}) the
Higgs fields develop vacuum expectation values 
\be
<\Phi_1>=\ds\frac{v_1}{\sqrt{2}}\,,\qquad\qquad<\Phi_2>=\ds\frac{v_2}{\sqrt{2}}
\label{41} 
\ee 
breaking the $SU(2)\times U(1)$ gauge symmetry and
generating masses for the bosons and fermions. The overall Higgs
norm $<\Phi>=\sqrt{v_1^2+v_2^2}=v=246\,\mbox{GeV}$ is fixed by the
Fermi constant. At the same time the ratio of the Higgs vacuum
expectation values remains arbitrary. Hence it is convenient to
introduce $\tan\beta=v_2/v_1$.

In general the interactions of the Higgs doublets $H_1$ and $H_2$
with quarks and leptons result in non--diagonal flavour
transitions \cite{19}. In particular these interactions contribute
to the amplitude of $K^0-\overline{K}^0$ oscillations and give
rise to new channels of muon decay like $\mu\to e^{-}e^{+}e^{-}$.
The common way to suppress flavour changing processes is to impose
a certain discrete $Z_2$ symmetry that forbids potentially
dangerous couplings of the Higgs fields to quarks and leptons
\cite{19}. Phenomenologically viable two--Higgs doublet models
obtained in such a way are classified according to the
interactions of $H_1$ and $H_2$ with fermions. Our initial
motivation encourages us to focus on the Higgs--fermion couplings
inherited from the minimal supersymmetric standard model, which
correspond to the Model II two Higgs doublet extension of the SM.
The Lagrangian of the 2HDM of type II is invariant under the
following symmetry transformations\footnote{Here $d_{Ri}$ and
$e_{Ri}$ denote the right-handed down-type quark and lepton
fields.}: 
\be 
H_1\to-H_1\,,\qquad d_{Ri}\to -d_{Ri}\,,\qquad
e_{Ri}\to -e_{Ri}\,, 
\label{411} 
\ee 
which forbid the couplings
$\lambda_6$ and $\lambda_7$ in the Higgs boson potential
(\ref{4}). The discrete symmetry (\ref{411}) also requires
$m_3^2=0$. But usually a soft violation of the symmetry
(\ref{411}) by dimension--two terms is allowed, since it does not
lead to Higgs--mediated tree--level flavor changing neutral
currents. Henceforth we set $\lambda_6=\lambda_7=0$ but retain a
non-vanishing value for $m_3^2$.

The invariance under the symmetry transformations (\ref{411})
ensures that only one Higgs doublet ($H_1$) interacts with the
down--type quarks and leptons, whereas the second one couples only
to up--type quarks \cite{211},\cite{16}. As a result, the running
masses of the $t$--quark ($m_t$), $b$-quark ($m_b$) and
$\tau$--lepton ($m_{\tau}$) in the 2HDM of type II are given by
\be
\ba{c} 
m_t(M_t)=\ds\frac{h_t(M_t) v}{\sqrt{2}}\sin\beta\, ,\qquad\qquad
m_b(M_t)=\ds\frac{h_b(M_t) v}{\sqrt{2}}\cos\beta\, , \\[3mm]
m_{\tau}(M_t)=\ds\frac{h_{\tau}(M_t) v}{\sqrt{2}}\cos\beta\, , 
\ea
\label{42} 
\ee 
where $M_t$ is the top quark pole mass. Since the
running masses of the fermions of the third generation are known,
Eq.(\ref{42}) is used to derive the Yukawa couplings $h_t(M_t)$,
$h_b(M_t)$ and $h_{\tau}(M_t)$, which play a crucial role in the
2HDM renormalization group flow.

Let us consider possible sets of global minima of the scalar
potential of the 2HDM of type II with vanishing energy density at
a high scale $\Phi \sim \Lambda$ and thereby degenerate with the
electroweak scale vacuum. If we ignore the running of the Higgs
self--couplings around this MPP scale $\Lambda$, then the most
favourable situation occurs when 
\be
\lambda_1(\Lambda)=\lambda_2(\Lambda)=\lambda_3(\Lambda)=
\lambda_4(\Lambda)=\lambda_5(\Lambda)=0\,.
\label{5} 
\ee 
In this case, for any vacuum configuration 
\be
<H_1>=\Phi_1\left( 
\ba{c}
0\\[2mm]
1
\ea
\right)\,,\qquad
<H_2>=\Phi_2\left(
\ba{c}
\sin\theta\\[2mm]
\cos\theta\, e^{i\omega}
\ea
\right)\,,\\[2mm]
\label{6} 
\ee 
where $\Phi_1^2+\Phi_2^2=\Lambda^2$, the quartic
part of the effective potential (\ref{4}) goes to zero. Here, the
gauge is fixed so that only the real part of the lower component
of $H_1$ gets a vacuum expectation value \footnote{The $U(1)$
gauge invariance allows us to eliminate the imaginary part of the
top component of $H_2$ as well.}.

But the 2HDM renormalization group flow then leads to the
instability of the vacua (\ref{6}). In fact for moderate values of
$\tan\beta$ the Higgs self--coupling $\lambda_1$ becomes negative
just below the MPP scale (see Fig.1a). The renormalization group
running of $\lambda_2$ exhibits the opposite behaviour, because of
the large and negative top quark contribution to the corresponding
beta--function. This means that, near the MPP scale, there is a
minimum with a huge and negative energy density $(\sim
-\Lambda^4)$ where $ <\Phi_2>=0$ and $<\Phi_1>\lesssim \Lambda$.

The renormalization group flow of $\lambda_1$ only changes at very
large $\tan\beta$ (see Fig.1b). The absolute value of the
$b$--quark and $\tau$--lepton contribution to $\beta_{\lambda_1}$,
being negligible at the moderate values of $\tan\beta$, grows with
increasing $\tan\beta$. At $\tan\beta\sim m_t(M_t)/m_b(M_t)$ their
negative contribution to the beta--function of $\lambda_1$
prevails over the positive contributions coming from the loops
containing Higgs and gauge bosons. The negative sign of
$\beta_{\lambda_1}$ results in $\lambda_1(\Phi)>0$ if the overall
Higgs norm $\Phi$ is less than $\Lambda$.

However the positive sign of $\lambda_1$ does not ensure the
stability of the vacua (\ref{6}). Substituting the vacuum
configuration (\ref{6}) into the quartic part of the 2HDM scalar
potential and omitting all bilinear terms in the Higgs fields one
finds for any $\Phi$ below the MPP scale: 
\be 
\ba{rcl}
V(H_1,H_2)&\approx&\ds\frac{1}{2}\biggl(\sqrt{\lambda_1(\Phi)}\Phi_1^2-
\sqrt{\lambda_2(\Phi)}\Phi_2^2\biggr)^2+\\[3mm]
&&+\left(\sqrt{\lambda_1(\Phi)\lambda_2(\Phi)}+\lambda_3(\Phi)+
\lambda_4(\Phi)\cos^2\theta\right) \Phi_1^2\Phi_2^2\, , 
\ea
\label{8} 
\ee 
Since the Higgs self--coupling $\lambda_5$ is taken
to be zero at the scale $\Lambda$, it is not generated at any
scale due to the form of the 2HDM renormalization group equations
\cite{171}--\cite{17}. The Higgs scalar potential (\ref{8})
attains its minimal value for $\cos\theta=0$ if $\lambda_4>0$ or
$\cos\theta=\pm 1$ when $\lambda_4<0$. Around the minimum the
scalar potential can be written as 
\be
V(H_1,H_2)\approx\ds\frac{1}{2}\biggl(\sqrt{\lambda_1(\Phi)}\Phi_1^2-
\sqrt{\lambda_2(\Phi)}\Phi_2^2\biggr)^2+\tilde{\lambda}(\Phi)\Phi_1^2\Phi_2^2\,,
\label{9} 
\ee 
where
$$
\tilde{\lambda}(\Phi)=\sqrt{\lambda_1(\Phi)\lambda_2(\Phi)}+\lambda_3(\Phi)+
\min\{0,\lambda_4(\Phi)\}\,.
$$
If at some intermediate scale the combination of the Higgs
self--couplings $\tilde{\lambda}(\Phi)$ is less than zero, then
there exists a minimum with negative energy density causing the
instability of the vacua at the electroweak and MPP scales.
Otherwise the Higgs effective potential is positive definite and
the considered vacua are stable.

In Fig.1b the Higgs self--couplings $\lambda_1(\Phi)$ and
$\lambda_2(\Phi)$ as well as the combination
$\tilde{\lambda}(\Phi)$ are plotted as a function of $\Phi$ for a
large value of $\tan\beta$. It is clear that the vacuum stability
conditions, i.e. 
\be 
\lambda_1(\Phi)\gtrsim
0\,,\qquad\qquad\lambda_2(\Phi)\gtrsim
0\,,\qquad\qquad\tilde{\lambda}(\Phi) \gtrsim 0 
\label{10} 
\ee 
are not fulfilled simultaneously. The value of $\tilde{\lambda}(\Phi)$
tends to be negative for $\Phi<\Lambda$. So the above
considerations demonstrate the failure of the original assumption
(\ref{5}), which therefore can not provide a self--consistent
realization of the MPP in the 2HDM.

\section{MPP conditions}

At the next stage it is worth relaxing the conditions (\ref{5})
and permitting $\lambda_1(\Lambda)$ and $\lambda_2(\Lambda)$ to
take on non--zero values \footnote{This assumption does not look
artificial if we take into account that the corresponding Higgs
self--couplings differ from zero in any phenomenologically
acceptable SUSY extension of the SM.}. Again the Higgs
self-coupling $\lambda_5(\Phi)$ remains zero at all scales. In
order to avoid a huge and negative vacuum energy density in the
global minimum of the 2HDM type II effective potential, the vacuum
stability conditions (\ref{10}) should be satisfied for any $\Phi$
in the interval: $v\lesssim \Phi\lesssim \Lambda$. In this case
both terms in the quartic part of the scalar potential (\ref{9})
are positive. In order to achieve the degeneracy of the vacua at
the electroweak and MPP scales, they must go to zero separately at
the scale $\Lambda$. For finite values of $\lambda_1(\Lambda)$ and
$\lambda_2(\Lambda)$ the first term in the quartic part of the
scalar potential (\ref{9}) can be eliminated by the appropriate
choice of Higgs vacuum expectation values 
\be
\Phi_1=\Lambda\cos\gamma\,,\qquad
\Phi_2=\Lambda\sin\gamma\,,\qquad
\tan\gamma=\Biggl(\ds\frac{\lambda_1}{\lambda_2}\Biggr)^{1/4}\,,
\label{12} 
\ee 
at which $V(H_1,H_2)$ attains its minimal value if
the vacuum stability conditions (\ref{10}) are fulfilled. The
vanishing of the second term in Eq.(\ref{9}) requires a certain
fine-tuning of the Higgs self--couplings at the MPP scale, namely
$\tilde{\lambda}(\Lambda)=0$. In order to get
$\tilde{\lambda}(\Lambda)=0$ at least one other Higgs
self--coupling, $\lambda_3(\Lambda)$ or $\lambda_4(\Lambda)$, has
to take on a non--zero value at the MPP scale. If the fine-tuning
between the Higgs self--couplings mentioned above takes place,
then the Higgs scalar potential (\ref{9}) tends to zero at the MPP
scale independently of the phase $\omega$ in the vacuum
configuration (\ref{6}).

At the first glance of Eq.(\ref{8}), it even appears possible to
get a set of degenerate vacua in which the energy density vanishes
for any value of angle $\theta$. This should correspond to 
\be
\lambda_3(\Lambda)=-\sqrt{\lambda_1(\Lambda)\lambda_2(\Lambda)}\,,\qquad
\lambda_4(\Lambda)=0\,. 
\label{11} 
\ee 
Nevertheless the situation
is not as promising as it first appears. The stability of the
vacuum configuration (\ref{6}) requires that the Higgs effective
potential does not go to negative values in close vicinity to the
MPP scale for $\Phi\gtrsim\Lambda$. In other words at the scale
$\Lambda$ there has to be a local minimum in which all partial
derivatives of the 2HDM scalar potential go to zero. The
degeneracy of the vacua at the MPP scale implies that they should
vanish for any choice of $\theta$ and $\omega$. Near the vacuum
configuration parameterized by Eq.(\ref{6}) and Eq.(\ref{12}) the
derivatives of $V(H_1,H_2)$ are proportional to 
\be
\ds\frac{\partial V(H_1,H_2)}{\partial \Phi_i}\propto
\ds\frac{1}{2}\beta_{\lambda_1}\tan^{-2}\gamma+
\frac{1}{2}\beta_{\lambda_2}\tan^2\gamma+\beta_{\lambda_3}+
\beta_{\lambda_4}\cos^2\theta\,.
\label{13} 
\ee 
These partial derivatives tend to zero for any
angles $\theta$ when $\beta_{\lambda_4}=0$. However for
$\lambda_4(\Lambda)=0$ the beta--function of $\lambda_4$ at the
scale $\Lambda$ is given by 
\be
\beta_{\lambda_4}=\ds\frac{1}{(4\pi)^2}\biggl[3g_2^2(\Lambda)g_1^2(\Lambda)
+12h_t^2(\Lambda)h_b^2(\Lambda)\biggr] 
\label{14} 
\ee 
where $g_2$ and $g_1$ are the $SU(2)$ and $U(1)$ gauge coupling constants. 
It is always positive and thereby spoils the stability of the vacua
given by Eq.(\ref{6}) and Eq(\ref{12}). Thus our attempt to adapt
the MPP idea to the 2HDM with $\lambda_4(\Lambda)=0$ fails.

For non--zero values of $\lambda_1(\Lambda)$ and
$\lambda_2(\Lambda)$ the self--consistent implementation of the
MPP can only be obtained if $\lambda_3(\Lambda)\ne 0$ and
$\lambda_4(\Lambda)\ne 0$. In order to ensure the degeneracy of
the physical and the MPP scale vacua and to satisfy the vacuum
stability constraints (\ref{10}), the combination of the Higgs
self--couplings $\tilde{\lambda}(\Lambda)$ and its derivative must
vanish simultaneously at the scale $\Lambda$. Then the 2HDM
effective potential possesses a set of local minima: 
\be
<H_1>=\left(
\begin{array}{c}
0\\ \Phi_1
\end{array}
\right)\; , \qquad <H_2>=\left(
\begin{array}{c}
\Phi_2\\ 0
\end{array}
\right)
\label{15}
\ee
when $\lambda_4(\Lambda)>0$ and
\be
<H_1>=\left(
\begin{array}{c}
0\\ \Phi_1
\end{array}
\right)\; , \qquad <H_2>=\left(
\begin{array}{c}
0\\ \Phi_2\, e^{i\omega}
\end{array}
\right) 
\label{16} 
\ee 
if $\lambda_4(\Lambda)$ is less than zero,
in which the vacuum energy density tends to zero. The Higgs field
norms $\Phi_1$ and $\Phi_2$ in the vacuum configurations
(\ref{15})--(\ref{16}) are determined by the equations for the
extrema of the 2HDM scalar potential, whose solution is given by
Eq.(\ref{12}). We should notice here that the existence of the
minimum (\ref{15}) does not necessarily require the vanishing of
$\lambda_5(\Lambda)$. Similar vacua with vanishing energy density
can also be obtained for non--zero values of this Higgs
self--coupling, if it satisfies the constraint:
$|\lambda_5(\Lambda)|<\lambda_4(\Lambda)$. At the minimum
(\ref{15}) the $SU(2)\times U(1)$ symmetry is broken completely
and the photon gains a mass of the order of $\Lambda$. Although
this is not in conflict with phenomenology, since an MPP scale
minimum is not presently realised in Nature, the scenario with
$\lambda_4(\Lambda)<0$ is more in compliance with the MPP
philosophy, simply because it results in a larger set of
degenerate vacua. In the minima (\ref{16}) the photon remains
massless and electric charge is conserved.

From the above considerations it becomes clear that the vacuum
configurations (\ref{16}) represent the largest possible set of
local degenerate minima of the Higgs effective potential, which
can be obtained in the 2HDM at the MPP high energy scale $\Lambda$
for non--zero values of $\lambda_1(\Lambda)$ and $\lambda_2(\Lambda)$.
The constraint on $\lambda_4(\Lambda)$ and the relationships
between different Higgs self--couplings 
\be 
\left\{ \ba{l}
\lambda_5(\Lambda)=0\,,\qquad \lambda_4(\Lambda)<0\\[2mm]
\tilde{\lambda}(\Lambda)=
\ds\frac{d\tilde{\lambda}(\Phi)}{d\Phi}\Biggl|_{\Phi=\Lambda}=0 \,, 
\ea 
\right. 
\label{17} 
\ee 
leading to the appearance of the
degenerate vacua, should be identified with the MPP conditions.
The conditions (\ref{17}) have to be supplemented by the vacuum
stability requirements (\ref{10}), which must be valid everywhere
from the electroweak scale to the MPP scale. Any failure of either the
conditions (\ref{17}) or the inequalities (\ref{10}) prevents the
consistent realization of the MPP in the 2HDM, when
$\lambda_1(\Lambda)\ne 0$ and $\lambda_2(\Lambda)\ne 0$.

Differentiating $\tilde{\lambda}$ near the MPP scale, replacing
the derivatives $\lambda_i'(\Lambda)$ by the corresponding
one--loop beta--functions and setting $\tilde{\lambda}'(\Lambda)$
to zero, one obtains two relations between the gauge, Yukawa and
Higgs self--couplings coupling constants at the MPP scale: 
\be
\lambda_3(\Lambda)=-\sqrt{\lambda_1(\Lambda)\lambda_2(\Lambda)}-
\lambda_4(\Lambda)\,,
\label{22} 
\ee 
\be 
\ba{c}
\lambda_4^2(\Lambda)=\ds\frac{6h_t^4(\Lambda)\lambda_1(\Lambda)}
{(\sqrt{\lambda_1(\Lambda)}+\sqrt{\lambda_2(\Lambda)})^2}+
\frac{(6h_b^4(\Lambda)+2h_{\tau}^4(\Lambda))\lambda_2(\Lambda)}
{(\sqrt{\lambda_1(\Lambda)}+\sqrt{\lambda_2(\Lambda)})^2}\,-\\[4mm]
\ds-2\lambda_1(\Lambda)\lambda_2(\Lambda)-\frac{3}{8}\biggl(3g_2^4(\Lambda)+
2g_2^2(\Lambda)g_1^2(\Lambda)+g_1^4(\Lambda)\biggr)\,. 
\ea 
\label{23} 
\ee 
The first of them
follows from $\tilde{\lambda}(\Lambda)=0$, whereas the second one
comes from the vanishing of the derivative of
$\tilde{\lambda}(\Phi)$ near the MPP scale. 
We note that, in the minimal SUSY
model, the MPP conditions (\ref{17}) are satisfied identically at
any scale lying higher than the superparticle masses. 

\section{Higgs spectrum}

Keeping in mind that Eq.(\ref{22})--(\ref{23}) relate different
couplings at the scale $\Lambda$ we can now explore the Higgs
spectrum in the vicinity of the physical vacuum of the MPP
inspired 2HDM of type II. The Higgs sector of the two Higgs
doublet extension of the SM involves eight states. Two linear
combinations of $\chi^{\pm}_1$ and $\chi^{\pm}_2$ are absorbed by
the $W^{\pm}$ bosons after the spontaneous $SU(2)\times U(1)$
symmetry breaking at the electroweak scale. A linear combination
of $A_1$ and $A_2$ become the longitudinal component of the $Z$
boson. The others form two charged and three neutral scalar
fields. One of the neutral Higgs bosons is CP--odd. The charged 
and CP-odd scalars do not interfere with each other and the CP-even 
states, because of the electric charge conservation and CP--invariance. 
They gain masses
\be
m^2_{\chi^{\pm}}=m_A^2-\ds\frac{\lambda_4}{2}v^2\,,\qquad\qquad
m_A^2=\frac{2m_3^2}{\sin 2\beta}\,, 
\label{25} 
\ee 
where $m_{\chi^{\pm}}$ and $m_A$ are the masses of the charged and
pseudoscalar Higgs bosons. The direct searches for the rare
B--meson decays ($B\to X_s\gamma$) place a lower limit on the
charged Higgs scalar mass in the 2HDM of type II \cite{24}: 
\be
m_{\chi^{\pm}}> 350\,\mbox{GeV}\,, 
\label{26} 
\ee 
which is also valid in our case.

In the basis
\be
\ba{rcl}
h_1&=&H_1^0\cos\beta+H_2^0\sin\beta\, ,\\
h_2&=&-H_1^0\sin\beta+H_2^0\cos\beta\,
\ea
\ee
the mass matrix of the CP--even Higgs fields is expressed as (see \cite{25})
\be
M^2=\left(
\ba{ll}
M^2_{11} & M^2_{12}\\[5mm]
M^2_{21} & M^2_{22}
\ea
\right)=
\left(
\ba{ll}
\frac{\ds\partial^2V}{\ds\partial \upsilon^2} \qquad~~~
\frac{\ds 1}{\ds \upsilon}\frac{\ds\partial^2V}{\ds\partial \upsilon\partial\beta}\\[5mm]
\frac{\ds 1}{\ds \upsilon}\frac{\ds\partial^2V}{\ds \partial \upsilon\partial\beta} \qquad
\frac{\ds 1}{\ds \upsilon^2}\frac{\ds \partial^2V}{\ds \partial\beta^2}
\ea
\right)\,,
\label{27}
\ee
$$
\ba{rcl}
M_{11}^2&=&\biggl(\lambda_1\cos^4\beta+\lambda_2\sin^4\beta+\ds\frac{\lambda}{2}\sin^22\beta\biggr)v^2\,,\\[2mm]
M_{12}^2&=&M_{21}^2=\ds\frac{v^2}{2}\biggl(-\lambda_1\cos^2\beta+\lambda_2\sin^2\beta+\lambda\cos 2\beta
\biggr)\sin 2\beta\,,\\[2mm]
M_{22}^2&=&m_A^2+\ds\frac{v^2}{4}\biggl(\lambda_1+\lambda_2-2\lambda\biggr)\sin^22\beta\,,
\ea
$$
where $\lambda=\lambda_3+\lambda_4$. Equations for the extrema of
the Higgs boson effective potential are used to eliminate $m_1^2$
and $m_2^2$ from Eq.(\ref{25}) and Eq.(\ref{27}). The top--left
entry of the CP--even mass matrix provides an upper bound on the
lightest Higgs scalar mass--squared. The masses of the two
CP--even eigenstates obtained by diagonalizing the matrix
(\ref{27}) are given by 
\be
m_{H,\,\,h}^2=\frac{1}{2}\left(M^2_{11}+M^2_{22}\pm\sqrt{(M_{22}^2-M_{11}^2)^2+4M^4_{12}}\right)\,.
\label{28} 
\ee 
With increasing $m_A$ the lightest Higgs boson mass
grows and approaches its maximum value $\sqrt{M_{11}^2}$ for
$m_A^2>>v^2$.

As follows from Eq.(\ref{25}) and Eq.(\ref{27}), the spectrum of
the Higgs bosons in the 2HDM of type II supplemented by the MPP
assumption is parameterized in terms of $m_A$, $\tan\beta$ and
four Higgs self--couplings $\lambda_1,\,\lambda_2,\,\lambda_3$ and
$\lambda_4$. Three other Higgs self--couplings $\lambda_5,
\lambda_6$ and $\lambda_7$ vanish due to the MPP conditions
(\ref{17}). At the scale $\Lambda$ the couplings
$\lambda_3(\Lambda)$ and $\lambda_4(\Lambda)$ can be expressed in
terms of $\lambda_1(\Lambda)$, $\lambda_2(\Lambda)$,
$g_i(\Lambda)$ and $h_j(\Lambda)$. The gauge couplings at the MPP
scale are fixed by $g_i(M_Z)$, which are extracted from the
electroweak precision measurements. The running of the Yukawa
couplings is mainly determined by $\tan\beta$. Thus, for a given
scale $\Lambda$, the evolution of the Higgs couplings are governed
by $\lambda_1(\Lambda), \lambda_2(\Lambda)$ and $\tan\beta$.
Therefore the Higgs masses and couplings depend on five variables
\be 
\lambda_1(\Lambda)\,,\qquad \lambda_2(\Lambda)\,\qquad
\Lambda\,\qquad \tan\beta, \qquad m_A\,. 
\label{29} 
\ee 
It means that, owing to the MPP, the model suggested in this article has
fewer free parameters compared to the 2HDM with an exact or softly
broken $Z_2$ symmetry. Therefore it can be considered as the
minimal non--supersymmetric two Higgs doublet extension of the SM.

\section{Numerical analysis}

The constraints on the Higgs masses in the 2HDM with an unbroken
$Z_2$ symmetry (with $m_3^2 = 0$) have been examined in a number
of publications \cite{17}, \cite{26}--\cite{27}. An analysis
performed assuming vacuum stability and the applicability of
perturbation theory up to a high energy scale (e.g. the
unification scale) revealed that all the Higgs boson masses lie
below $200\,\mbox{GeV}$ \cite{27}. Stringent restrictions on
the masses of the charged and pseudoscalar states were found. They
do not exceed $150\,\mbox{GeV}$. This upper bound is considerably
less than the lower experimental limit on $m_{\chi^{\pm}}$
(\ref{26}) obtained in the 2HDM of type II. This shows that 2HDM
with unbroken $Z_2$ symmetry is inconsistent with experimental
data.

The aim of our study is to analyse the MPP scenario in the 2HDM of
type II ($m_3^2\ne 0$) and compare it with that in the SM. As part of the
analysis sufficiently large values of $\tan\beta$ should be taken.
The motivation for this is quite simple. The top quark Yukawa
coupling at the electroweak scale approaches its SM value for
$\tan\beta>>1$. If simultaneously $\tan\beta$ is much less than
$m_t(M_t)/m_b(M_t)$ the $b$--quark and $\tau$--lepton Yukawa
couplings remain small and can be disregarded. Since in the
considered limit the beta--functions of $h_t$ in the SM and 2HDM
coincide, the renormalization group flows of the top quark Yukawa
coupling in these models are then identical and the main
differences in the spectra are caused by the Higgs couplings.

For the numerical analysis we adopt the following procedure. At
the first stage we fix the values of $\tan\beta$ ($\tan\beta=10$)
and the MPP scale ($\Lambda=M_{Pl}$). Then using the 2HDM
renormalization group equations we calculate the gauge and Yukawa
couplings at the scale $\Lambda$. For each given set of
$\lambda_1(\Lambda)$ and $\lambda_2(\Lambda)$ we define
$\lambda_3(\Lambda)$ and $\lambda_4(\Lambda)$ in accordance with
Eq.(\ref{22})--(\ref{23}), evolve the renormalization group
equations down, determine the values of all Higgs self--couplings
at the electroweak scale ($\mu=175\,\mbox{GeV}$) and study the
Higgs spectrum as a function of the pseudoscalar mass. After that
we investigate the dependence of the Higgs masses on
$\lambda_1(\Lambda)$ and $\lambda_2(\Lambda)$. At the next stage
we vary the value of $\tan\beta$ from 2 to 50 and MPP scale within
the interval: $10\,\mbox{TeV}\lesssim\Lambda\lesssim M_{Pl}$.
Finally the sensitivity of the Higgs masses to the choice of
$\alpha_3(M_Z)$, $m_t(M_t)$ and renormalization scale $\mu$ is
examined.

The results of our numerical study are summarized in Tables 1--2
and Figs. 2--3. The MPP assumption constrains $\lambda_1(\Lambda)$
and $\lambda_2(\Lambda)$ very strongly at moderate and large
values of $\tan\beta$. In Fig.2 different curves restrict the
allowed range of the corresponding Higgs self--couplings where
$\lambda_4^2(\Lambda)\ge 0$. Outside the allowed range
$\lambda_4(\Lambda)$ is complex and the Higgs effective potential
(\ref{4}) is non-hermitian. If the MPP scale is situated at very
high energies (e.g. $\Lambda\simeq M_{Pl}$) then only extremely
small values of $\lambda_2(\Lambda)\simeq 0.01$ are permitted for
most of the large $\tan\beta$ region. The ratio of
$\lambda_1(\Lambda)$ to $\lambda_2(\Lambda)$ is also limited so
that $\lambda_1(\Lambda)>>\lambda_2(\Lambda)$ since $\sin^2\gamma$
is bounded from below (see Fig.2a and 2b). These restrictions are
substantially relaxed at $\tan\beta\sim 1$ and at very large
values of $\tan\beta$ close to the upper limit coming from the
validity of perturbation theory at high energies. When $\tan\beta$
is of the order of $m_t(M_t)/m_b(M_t)$ the Higgs self--coupling
$\lambda_2(\Lambda)$ can be even much larger than
$\lambda_1(\Lambda)$ as the low values of $\sin^2\gamma$ are not
ruled out by the MPP in this case (see Fig.3a).

The restrictions on the Higgs self--couplings arising out of the
MPP become weaker in the scenarios with a low MPP scale. The
allowed regions of $\lambda_1(\Lambda)$ and $\lambda_2(\Lambda)$
enlarge because of the increase in the top quark Yukawa coupling
contribution to the right hand side of Eq.(\ref{23}). As before,
the admissible ranges of $\lambda_1(\Lambda)$ and
$\lambda_2(\Lambda)$ expand at moderate and very large values of
$\tan\beta$ (see Fig.2c--2d and Fig.3b).

The applicability of perturbation theory up to the scale $\Lambda$
and the requirement of the stability of the degenerate vacua
constrain the Higgs self--couplings further. While
$\lambda_2(\Lambda)$ can vary from zero to its upper bound (see
Fig.2a and Fig.2c), the Higgs self--coupling $\lambda_1(\Lambda)$
is limited from below and above for most of the $\tan\beta$
region. When the values of $\lambda_1(\Lambda)$ are too large they
either violate perturbativity or make the term
$\lambda_1(\Lambda)\cdot\lambda_2(\Lambda)$ in Eq.(\ref{23}) so
large that $\lambda_4^2(\Lambda)$ tends to be negative. Values of
$\lambda_1(\Lambda)$ which are too small either reduce the top
quark Yukawa contribution to the right hand side of Eq.(\ref{23}),
so that $\lambda_4^2(\Lambda)$ turns out to be negative, or result
in the changing of the sign of $\tilde{\lambda}(\Phi)$ during the
renormalization group flow giving rise to vacuum instability. The
allowed intervals of $\lambda_1(M_{Pl})$ are indicated in Table 1
for $\tan\beta=10$ and $\lambda_2(M_{Pl})=0.005$, for
$\tan\beta=2$ and $\lambda_2(M_{Pl})=0.05$ and for $\tan\beta=50$
and $\lambda_2(M_{Pl})=0.01$. Also Table 1 shows the admissible
ranges of $\lambda_1(\Lambda)=\lambda_2(\Lambda)$ for
$\Lambda=10\,\mbox{TeV}$ and the three different values of
$\tan\beta$. One can see that the lower bound on
$\lambda_1(\Lambda)$ weakens at very large values of
$\tan\beta\simeq m_t(M_t)/m_b(M_t)$ and in the scenarios when the
MPP conditions (\ref{17}) are fulfilled at low energies.

The restrictions on the Higgs self--couplings discussed above and
the choice of $m_3^2$ needed to respect the lower limit on the
charged scalar mass (\ref{26}), deduced from the non-observation
of $B\to X_s\gamma$ decay, maintain a mass hierarchy in the Higgs
sector of the MPP inspired 2HDM. Indeed the MPP, in conjunction
with the requirements of vacuum stability and validity of
perturbation theory, keep $\lambda_i v^2$ and the CP--even mass
matrix elements $M_{11}^2$ and $M_{12}^2$ well below $v^2$ for a
wide set of $\tan\beta$ ($\tan\beta<< m_t(M_t)/m_b(M_t)$)
and $\Lambda\gtrsim 10\,\mbox{TeV}$. Then the pseudoscalar mass
has to be substantially larger than $v$ in order to suppress the
branching ratio $B\to X_s\gamma$. As a consequence the masses of
the heaviest scalar, pseudoscalar and charged Higgs bosons are
confined around $m_A$ while the lightest CP-even Higgs state has a
mass of the order of the electroweak scale, i.e.: 
\be 
m_h^2\simeq M_{11}^2-\ds\frac{M_{12}^4}{m_A^2}+O\left(
\ds\frac{\lambda_i^2 v^4}{m_A^4}\right) \,. 
\label{291} 
\ee 
The results of the
numerical analysis given in Table 1 indicate that, for a wide
range of $\tan\beta$ ($2\lesssim\tan\beta\ll 50$), the lightest
Higgs boson mass is less than $180\,\mbox{GeV}$ for any reasonable
choice of the scale $\Lambda\gtrsim 10\,\mbox{TeV}$. Furthermore,
because of the large splitting among the Higgs boson masses, the
lightest Higgs scalar has the same couplings to fermions, W and
Z--bosons as the Higgs particle in the SM.

In order to illustrate how the MPP requires a value for $m_h$
around the top quark mass, let us assume that the MPP conditions
(\ref{17}) are fulfilled at the electroweak scale and
$\lambda_1(\Lambda)=\lambda_2(\Lambda)=\lambda_0$. Then in the
decoupling limit, when $m_A>>\lambda_i v$, the lightest Higgs
scalar mass is given by 
\be 
m_h^2\simeq\lambda_0 v^2
\cos^22\beta\,. 
\label{30} 
\ee 
From Eq.(\ref{30}) and the results
of the numerical studies given in Table 1, it becomes clear that
the mass of the lightest Higgs particle grows with increasing
$\lambda_0$ and $\tan\beta$. The allowed range of the Higgs
self--couplings is constrained by the MPP assumption. In the
interval $1\ll \tan\beta\ll m_t(M_t)/m_b(M_t)$ the MPP constraint 
(\ref{23}) implies that
$\lambda_0^2\le\ds\frac{3}{4}\left(h_t^4-\frac{3}{4}g_2^4-
\frac{1}{2}g_2^2g_1^2-\frac{1}{4}g_1^4\right)$.
As a result we get an upper limit on the mass of the lightest
Higgs boson 
\be 
m_h^2\lesssim \sqrt{3\biggl(m_t^4-(2\cos^4\theta_W+1)M_Z^4\biggr)}\simeq\sqrt{3}m_t^2\,, 
\label{32} 
\ee
which is set by the top quark mass. Here $\theta_W$ is the Weinberg angle.

A remarkable prediction for the mass of the SM--like Higgs boson
appears if the MPP scale $\Lambda$ is taken to a very high energy.
For large $\tan\beta$, the admissible region of the Higgs
self--couplings shrinks drastically (see Fig.2a) and only small
enough values of $\lambda_2(\Lambda)$, $\lambda_3(\Lambda)$ and
$\lambda_4(\Lambda)$ are allowed. Therefore the running of
$\lambda_2$ resembles the renormalization group flow of $\lambda$
in the SM with $\lambda(\Lambda)=0$. Since, at large $\tan\beta$,
the lightest Higgs boson mass is predominantly determined by the
term proportional to $\lambda_2 v^2$ in the top--left entry of the
CP--even Higgs mass matrix (\ref{27}), $\lambda_1(\Lambda)$
affects $m_h$ only marginally and the MPP prediction for the Higgs
mass in the SM is almost reproduced when $\Lambda$ is close to the
Planck scale (see Table 1--2). The main ambiguity in the
calculation of the SM--like Higgs boson mass is related to the
uncertainty of the top quark mass measurements\footnote{Unlike in
the SM, the top quark mass is not predicted by MPP in the 2HDM of
type II}. When the running top quark mass changes from
$165\,\mbox{GeV}$ to $170\,\mbox{GeV}$ $m_h$ increases by
$10\,\mbox{GeV}$ (see Table 2). The mass of the lightest Higgs
scalar is less sensitive to the choice of the scale $\mu$ 
down to which the 2HDM renormalization group equations are 
assumed valid.
It leads to only
a $6\,\mbox{GeV}$ uncertainty. Ultimately, for $\Lambda=M_{Pl}$
and relatively large values of $\tan\beta$, we find: 
\be
m_h=137\pm 12\,\,\mbox{GeV}\,. 
\label{33} 
\ee

The range of variation in the mass of the SM--like Higgs boson
enlarges at moderate and very large values of $\tan\beta$. At
$\tan\beta\simeq m_t(M_t)/m_b(M_t)$ the mass of the lightest Higgs
scalar increases because the strict upper limit on
$\lambda_2(\Lambda)$ is loosened, due to the large contribution to
the beta--functions of the Higgs self--couplings coming from the
loops containing the $b$--quark and the $\tau$--lepton. Now the
prediction (\ref{33}) represents the lower bound on $m_h$ for
$\Lambda = M_{Pl}$, while the restriction on the lightest Higgs
scalar mass from above tends to the upper bound on $m_h$ in the SM
--- $180\,\mbox{GeV}$. At moderate values of $\tan\beta$ the mass
of the lightest Higgs particle decreases. As a result, at
$\tan\beta=2$, one can easily get $m_h=114\,\mbox{GeV}$ without
any modification of the MPP, such as that suggested for the SM in
\cite{29}.

With a lowering of the MPP scale, the allowed range of $m_h$
expands. At $\tan\beta=50$ and $\Lambda=10\,\mbox{TeV}$ the
SM--like Higgs boson mass can be even as heavy as
$300\,\mbox{GeV}$ if $\lambda_2(\Lambda)>>\lambda_1(\Lambda)$ (see
Table 1). However if $\tan\beta$ is not so large ($\tan\beta\ll
m_t(M_t)/m_b(M_t)$), $m_h$ remains lower than $180\,\mbox{GeV}$
because of the stringent limit on $\lambda_2(\Lambda)$ (see
Fig.2c). The upper bound on $m_h$ is even stronger for moderate
$\tan\beta$, where a considerable part of the parameter space is
excluded by the unsuccesful Higgs searches at LEP.

\section{Conclusions}

We have constructed a new minimal non--supersymmetric two Higgs
doublet extension of the SM by applying the MPP assumption to the
SUSY inspired 2HDM. According to the MPP, $\lambda_5$ vanishes,
preserving CP--invariance. Four other Higgs self--couplings obey
two relationships (\ref{17}) at some scale $\Lambda>>v$ leading to
the largest possible set of degenerate vacua in the SUSY inspired
2HDM. Usually the existence of a large set of degenerate vacua is
associated with an enlarged global symmetry of the Lagrangian. The
2HDM is not an exception. When $m_3^2$, $\lambda_5$, $\lambda_6$
and $\lambda_7$ in the Higgs effective potential (\ref{4}) vanish,
the Lagrangian of the 2HDM is invariant under the transformations
of an $SU(2)\times [U(1)]^2$ global symmetry. The additional
$U(1)$ symmetry is nothing else than the Peccei--Quinn symmetry
introduced to solve the strong CP problem in QCD \cite{21}. The
mixing term $m_3^2(H_1^{\dagger}H_2)$ in the effective potential
(\ref{4}), which is not forbidden by the MPP, breaks the extra
$U(1)$ global symmetry softly so that a pseudo-Goldstone boson
(the axion) does not appear in the particle spectrum. At high
energies, where the contribution of the mixing term
$m_3^2(H_1^{\dagger}H_2)$ can be safely ignored, the invariance under the
Peccei--Quinn symmetry is restored giving rise to the set of
degenerate vacua (\ref{16}). The MPP predictions for $\lambda_i$
(\ref{17})--(\ref{23}) can be tested when the masses and couplings
of the Higgs bosons are measured at the future colliders.

In the large $\tan\beta$ limit, when 2HDM approaches the SM, the
allowed range of the Higgs self--couplings is severely constrained
by the MPP conditions (\ref{17}) and vacuum stability requirements
(\ref{10}). As a consequence, for most of the large $\tan\beta$
($\tan\beta\gtrsim 2$) region the Higgs spectrum exhibits a
hierarchical structure. While the heavy scalar, pseudoscalar and
charged Higgs particles are nearly degenerate around $m_A$, and
the latter should be greater than $350\,\mbox{GeV}$ owing to the
stringent limit on the $m_{\chi^{\pm}}$ (\ref{26}) coming from the
searches of the rare B--meson decays, the mass of the SM--like
Higgs boson $m_h$ does not exceed $180\,\mbox{GeV}$ for any scale
$\Lambda\gtrsim 10\,\mbox{TeV}$. The theoretical bound on $m_h$
obtained here is quite stringent. For comparison the lightest
Higgs boson in the 2HDM with a softly broken $Z_2$ symmetry can be
even heavier than $400\,\mbox{GeV}$ \cite{28} for $\Lambda\simeq
10\,\mbox{TeV}$. So a fairly stringent constraint on $m_h$ arises
from the application of the MPP to the 2HDM of type II.

The bounds on $m_h$ become even stronger if the MPP conditions are
realized at high energies. In this case the MPP prediction for the
Higgs mass obtained in the SM is reproduced. But, in contrast to
the SM, lower values of $m_h\simeq 115\,\mbox{GeV}$ may be easily
obtained in the MPP inspired 2HDM when $\tan\beta$ approaches $2$.
The restrictions on the Higgs self--couplings and the mass of the
lightest Higgs particle are less stringent at very large values of
$\tan\beta\simeq m_t(M_t)/m_b(M_t)$.

\vspace{3mm}
\noindent
{\large \bf Acknowledgements}

\noindent The authors are grateful to S.King, M.Krawczyk, S.Moretti 
and M.Vysotsky for fruitful discussions. RN would like to acknowledge
support from the PPARC grant PPA/G/S/2003/00096. RN was also
partly supported by a Grant of the President of Russia for young
scientists (MK--3702.2004.2). The work of CF was supported by
PPARC and the Niels Bohr Institute Fund. CF would like to
acknowledge the hospitality of the Niels Bohr Institute.
LL thanks all participants at her Niels Bohr Institute theoretical 
seminar for fruitful discussions and the Russian Foundation for Basic 
Research (RFBR) for financial support, project 05-02-17642.

\newpage

\bc {\bf Table 1.} The lightest running Higgs mass for
$m_A=400\,\mbox{GeV}$, $M_t=175\,\mbox{GeV}$ and
$\alpha_3(M_Z)=0.117$
($m_h$ is given in GeV and calculated at the scale $\mu=175\,\mbox{GeV}$). \\
\vspace{5mm}
\normalsize{
\begin{tabular}{|c|c|c|c|c|}
\hline
$\Lambda$ & $\tan\beta$ & $\lambda_1(\Lambda)$ & $\lambda_2(\Lambda)$ & $m_h$\\
&&&&\\
\hline
&& 1.0 & 0.005 & 137.8 \\
\cline{3-5}
&& 3.5 & 0.005 & 137.9 \\
\cline{3-5}
&$\tan\beta=10$& 0.25 & 0.005 & 138.5 \\
\cline{3-5}
&& 1.0 & 0.008 & 138.2 \\
\cline{3-5}
&& 1.0 & 0.001 & 136.8 \\
\cline{2-5}
&& 1.6 & 0.05 & 118.1 \\
\cline{3-5}
&& 3.2 & 0.05 & 128.3 \\
\cline{3-5}
$\Lambda=M_{Pl}$ & $\tan\beta=2$ & 0.85 & 0.05 & 116.7\\
\cline{3-5}
&& 1.6 & 0.08 & 127.4 \\
\cline{3-5}
&& 1.6 & 0.02 & 114.9 \\
\cline{2-5}
&& 1.0 & 0.01 & 140.9 \\
\cline{3-5}
&& 3.0 & 0.01 & 141.6 \\
\cline{3-5}
& $\tan\beta=50$ & 0.1 & 0.01 & 141.8 \\
\cline{3-5}
&& 0.04 & 0.1 & 148.4 \\
\cline{3-5}
&& 0.01 & 4.0 & 170.7\\
\hline
&& 0.25 & 0.25 & 142.0\\
\cline{3-5}
&& 0.45 & 0.45 & 166.6\\
\cline{3-5}
& $\tan\beta=10$ & 0.10 & 0.10 & 115.3\\
\cline{3-5}
&& 0.25 & 0.45 & 168.2\\
\cline{3-5}
&& 2.4 & 0.25 & 134.7\\
\cline{2-5}
&& 0.3 & 0.3 & 103.2\\
\cline{3-5}
&& 0.65 & 0.65 & 116.6\\
\cline{3-5}
$\Lambda=10\,\mbox{TeV}$ & $\tan\beta=2$ & 0.16 & 0.16 & 95.6\\
\cline{3-5}
&& 0.3 & 0.7 & 131.5\\
\cline{3-5}
&& 4.0 & 0.3 & 72.4\\
\cline{2-5}
&& 0.3 & 0.3 & 150.2\\
\cline{3-5}
&& 0.64 & 0.64 & 188.9\\
\cline{3-5}
& $\tan\beta=50$ & 0.01 & 0.01 & 89.0 \\
\cline{3-5}
&& 0.1 & 4.0 & 321.9\\
\cline{3-5}
&& 4.0 & 0.1 & 114.6\\
\hline
\end{tabular}}
\ec

\newpage
\bc {\bf Table 2.} The Higgs spectrum for $\Lambda=M_{Pl}$,
$\tan\beta=10$, $\lambda_1(M_{Pl})=1$ and
$\lambda_2(M_{Pl})=0.005$ (all masses are given in GeV).\\
\vspace{5mm} \normalsize{
\begin{tabular}{|c|c|c|c|c|c|c|c|c|}
\hline
$m_A$ & 400 & 400 & 400 & 400 & 400 & 400 & 200 & 1000 \\
\hline
$m_t$ & 165 & 165 & 165 & 165 & 165 & 170 & 165 & 165\\
\hline
$\mu$& 175 & $M_Z$ & 400 & 175 & 175 & 180 & 175 & 175\\
\hline
$\alpha_{3}(M_Z)$& 0.117 & 0.117 & 0.117 & 0.119 & 0.115 & 0.117 & 0.117 & 0.117\\
\hline
$m_h(\mu)$& 137.8 & 143.3 & 131.4 & 137.0 & 138.6 & 146.6 & 136.7 & 137.9 \\
\hline
$m_H(\mu)$& 400.8 & 400.8 & 400.8 & 400.8 & 400.8 & 400.9 & 202.4 & 1000.3\\
\hline
$m_{\chi^{\pm}}(\mu)$& 406.7 & 406.7 & 406.7 & 405.2 & 407.8 & 411.0 & 213.1 & 1002.7\\
\hline
\end{tabular}}
\ec

\newpage
\noindent
{\Large \bf Figure captions}
\vspace{5mm}

\noindent {\bf Fig.1.}\,\,The running of $\lambda_1$, $\lambda_2$
and $\tilde{\lambda}$ below $M_{Pl}$ for $\lambda_i(M_{Pl})=0$,
$M_t=175\,\mbox{GeV}$ and $\alpha_3(M_Z)=0.117$ for (a)
$\tan\beta=2$ and (b) $\tan\beta=50$.  The solid, dashed and
dash--dotted lines correspond to $\lambda_1$, $\lambda_2$ and
$\tilde{\lambda}$ respectively. The running of $\tilde{\lambda}$
is not shown for $\tan\beta=2$ because
$\tilde{\lambda}$ becomes complex when $\lambda_1<0$.\\
\\
{\bf Fig.2.}\,\,The MPP bounds on (a) $\lambda_2(\Lambda)$ and (b)
$\lambda_1(\Lambda)$, for $\Lambda=M_{Pl}$, as a function of
$\sin^2\gamma$. The corresponding bounds on (c)
$\lambda_2(\Lambda)$ and (d) $\lambda_1(\Lambda)$ for
$\Lambda=10\,\mbox{TeV}$ are also shown. The solid and
dash--dotted curves represent the limits on the Higgs
self--couplings for $\tan\beta=10$ and $\tan\beta=2$ respectively.
The allowed range of the Higgs self--couplings lies below curves.\\
\\
{\bf Fig.3.}\,\,The MPP restrictions on
$\lambda_1(\Lambda)\cdot\lambda_2(\Lambda)$ for $\tan\beta = 50$
versus $\sin^2\gamma$ for (a) $\Lambda=M_{Pl}$ and (b)
$\Lambda=10\,\mbox{TeV}$. The allowed part of the parameter space
lies below the curves.

\newpage
{\large \hspace{-1.5cm}$\lambda_i(\Phi)$}
\begin{center}
{\hspace*{-20mm}\includegraphics[height=100mm,keepaspectratio=true]{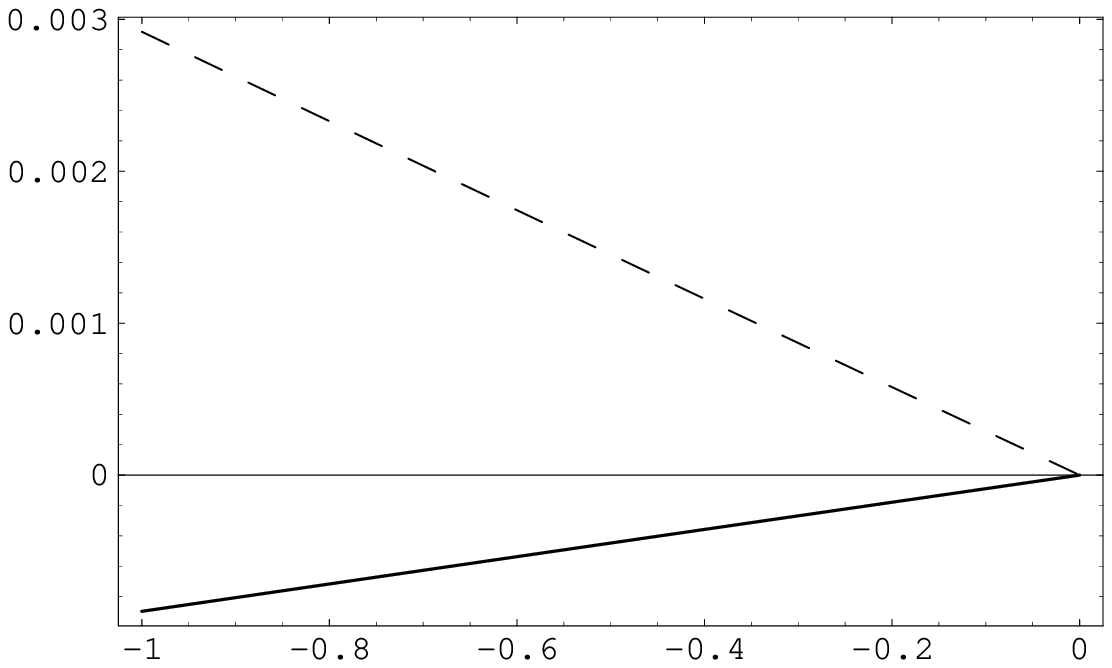}}\\
{\large $\log[\Phi^2/M_{Pl}^2]$}\\[0mm] {\large\bfseries Fig.1a}
\end{center}
{\large \hspace{-0.5cm}$\lambda_i(\Phi)$}
\begin{center}
{\hspace*{-20mm}\includegraphics[height=100mm,keepaspectratio=true]{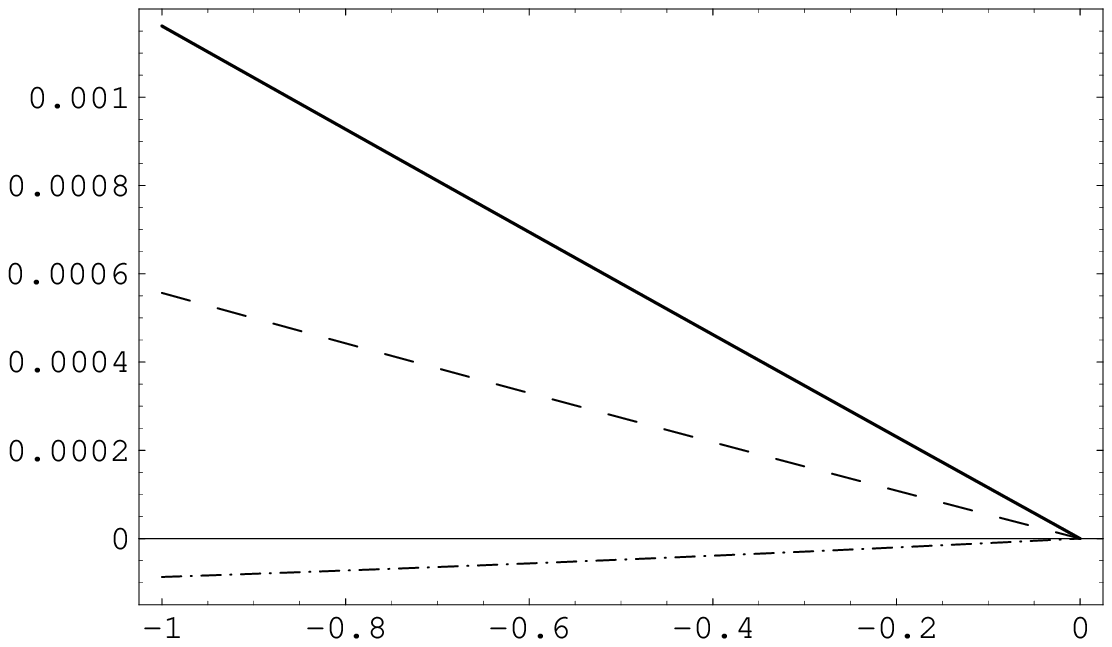}}\\
{\large $\log[\Phi^2/M_{Pl}^2]$}\\[0mm] {\large\bfseries Fig.1b}
\end{center}
\vspace{-24cm}
{\large
\hspace{12cm}
$
\ba{l}
\tan\beta=2\\[12.1cm]
\tan\beta=50
\ea
$}\\

\newpage
{\large \hspace{-1.5cm}$\lambda_2(M_{Pl})$}
\begin{center}
{\hspace*{-20mm}\includegraphics[height=100mm,keepaspectratio=true]{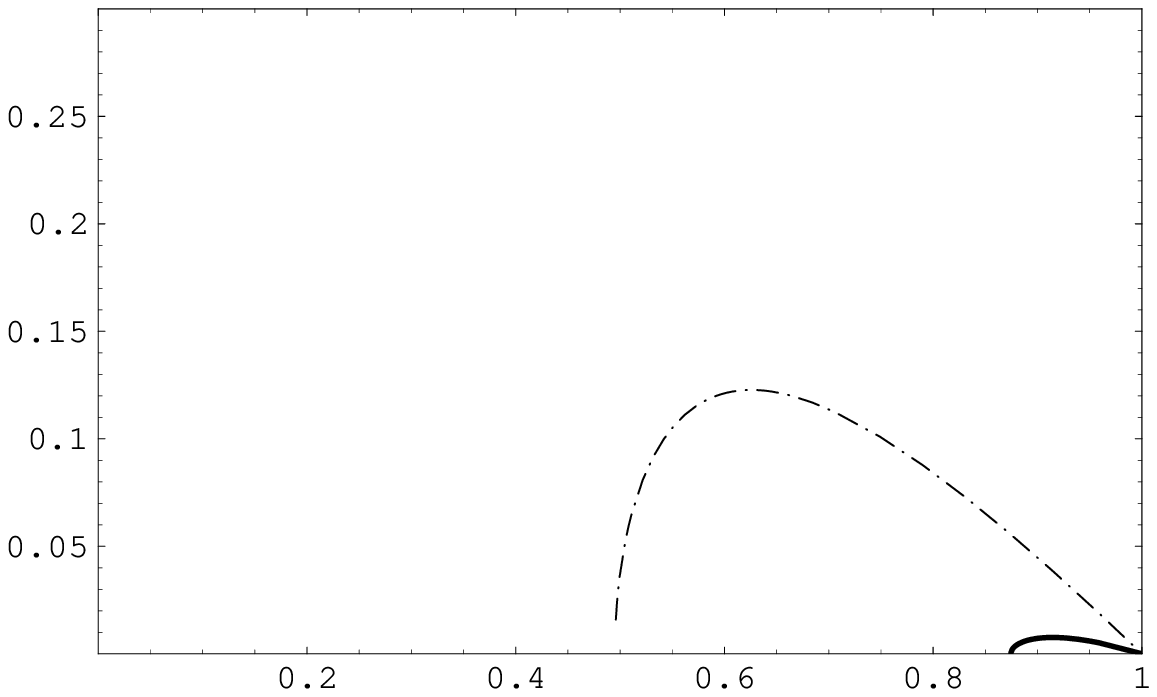}}\\
{\large $\sin^2\gamma$}\\ {\large\bfseries Fig.2a}\\
\end{center}
{\large \hspace{-1cm}$\lambda_1(M_{Pl})$}
\begin{center}
{\hspace*{-20mm}\includegraphics[height=100mm,keepaspectratio=true]{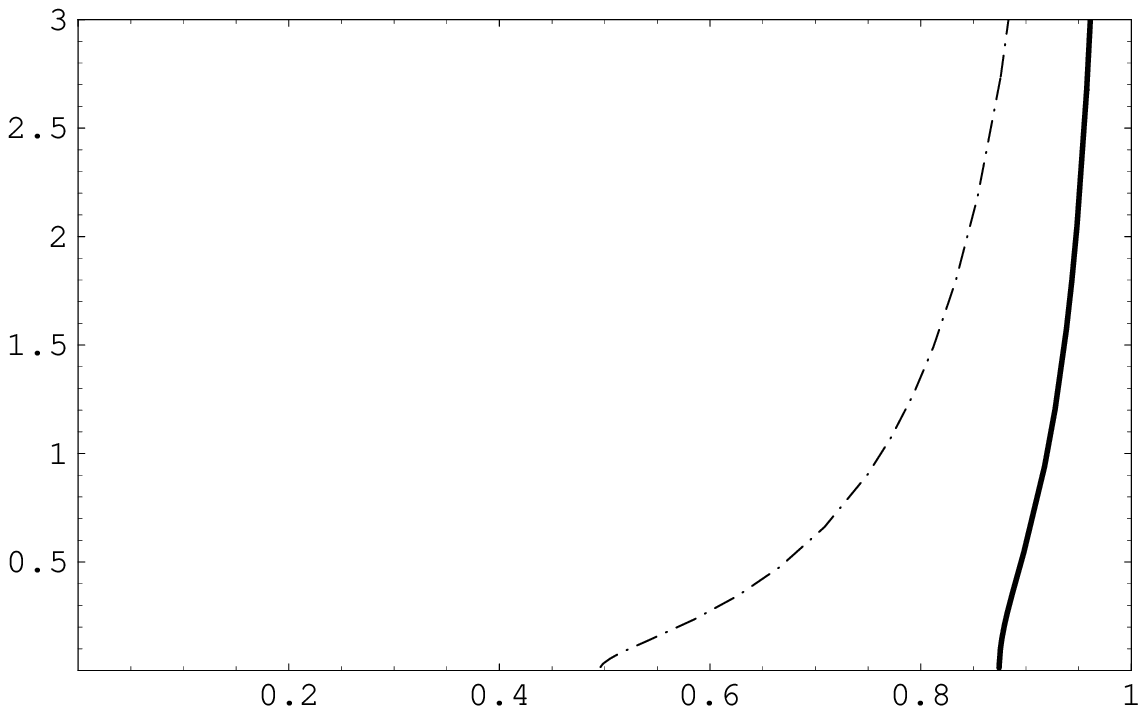}}\\
{\large $\sin^2\gamma$}\\ {\large\bfseries Fig.2b}
\end{center}

\newpage
{\large \hspace{-1.5cm}$\lambda_2(\Lambda)$}
\begin{center}
{\hspace*{-20mm}\includegraphics[height=100mm,keepaspectratio=true]{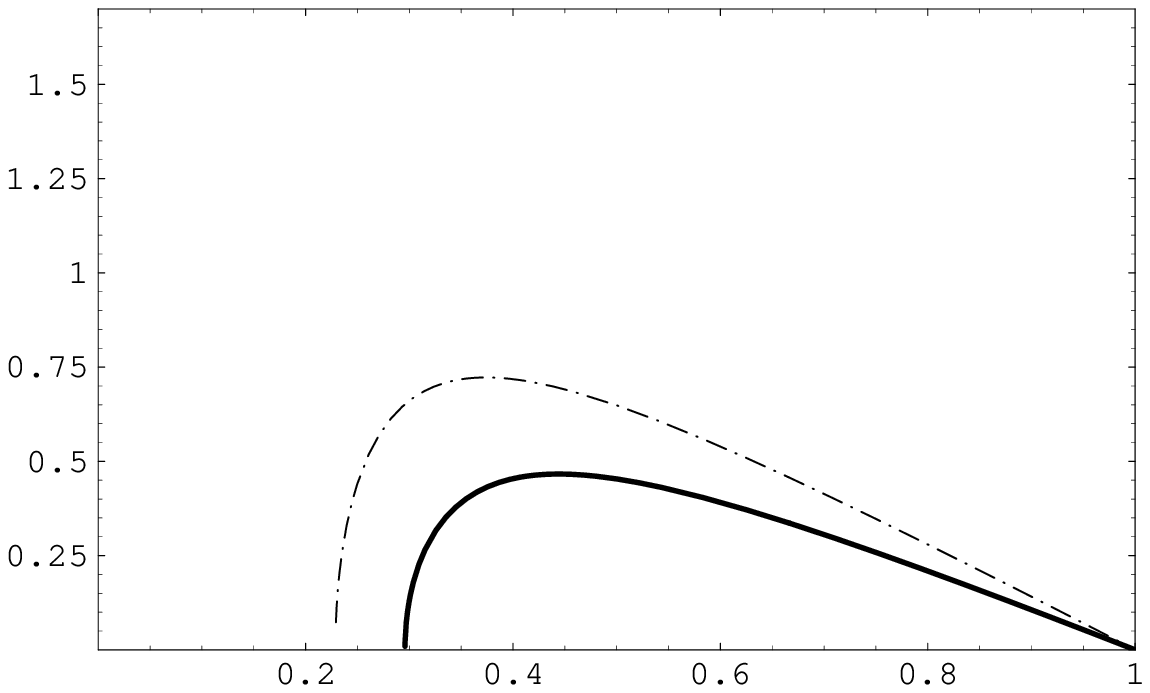}}\\
{\large $\sin^2\gamma$}\\ {\large\bfseries Fig.2c}\\
\end{center}
{\large \hspace{-1cm}$\lambda_1(\Lambda)$}
\begin{center}
{\hspace*{-20mm}\includegraphics[height=100mm,keepaspectratio=true]{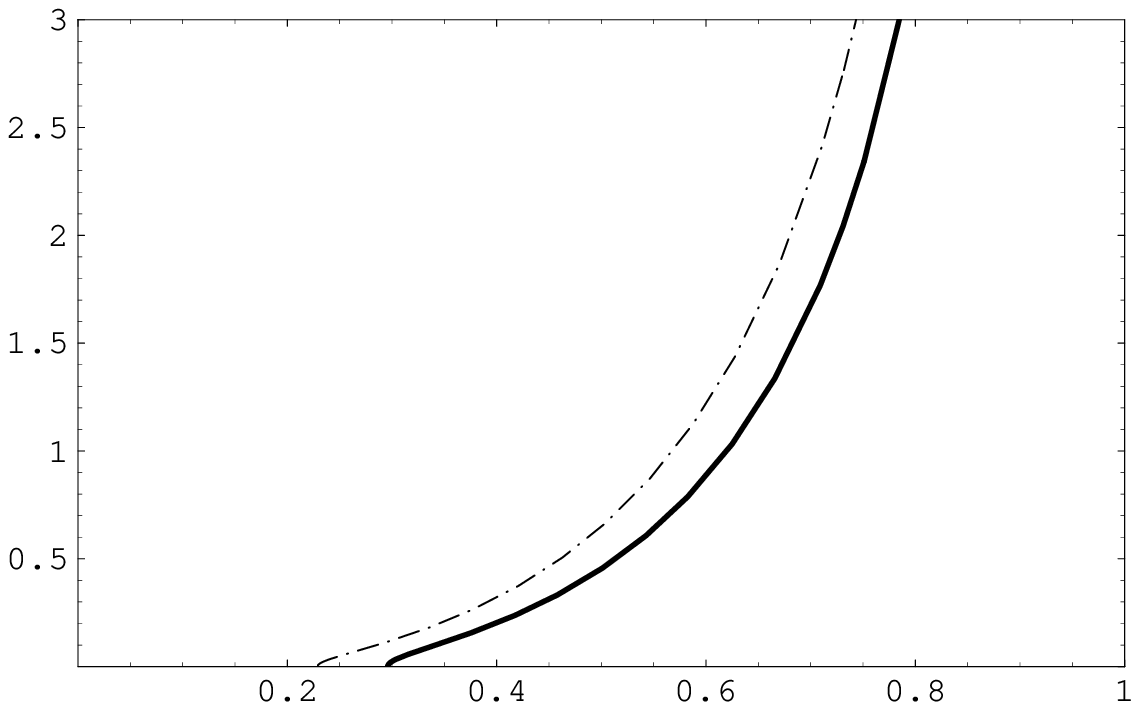}}\\
{\large $\sin^2\gamma$}\\ {\large\bfseries Fig.2d}
\end{center}

\newpage
{\large \hspace{-2cm}$\lambda_2(M_{Pl})\cdot\lambda_1(M_{Pl})$}
\begin{center}
{\hspace*{-20mm}\includegraphics[height=100mm,keepaspectratio=true]{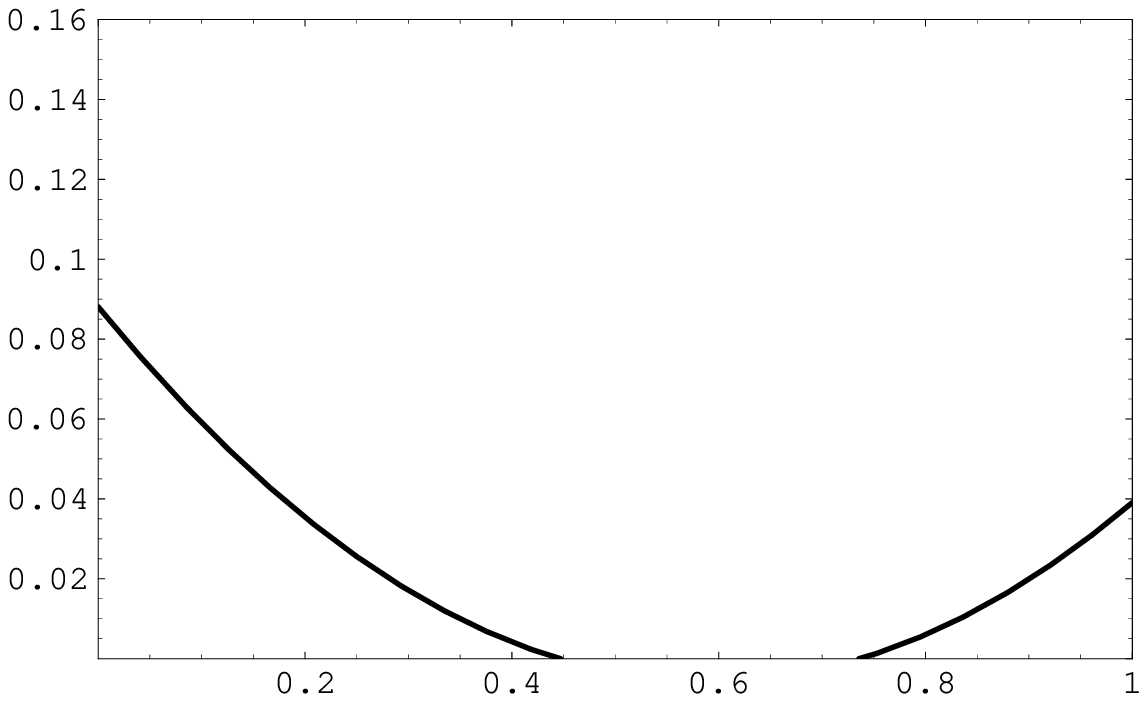}}\\
{\large $\sin^2\gamma$}\\
{\large\bfseries Fig.3a}\\
\end{center}
{\large \hspace{-1.5cm}$\lambda_2(\Lambda)\cdot\lambda_1(\Lambda)$}
\begin{center}
{\hspace*{-20mm}\includegraphics[height=100mm,keepaspectratio=true]{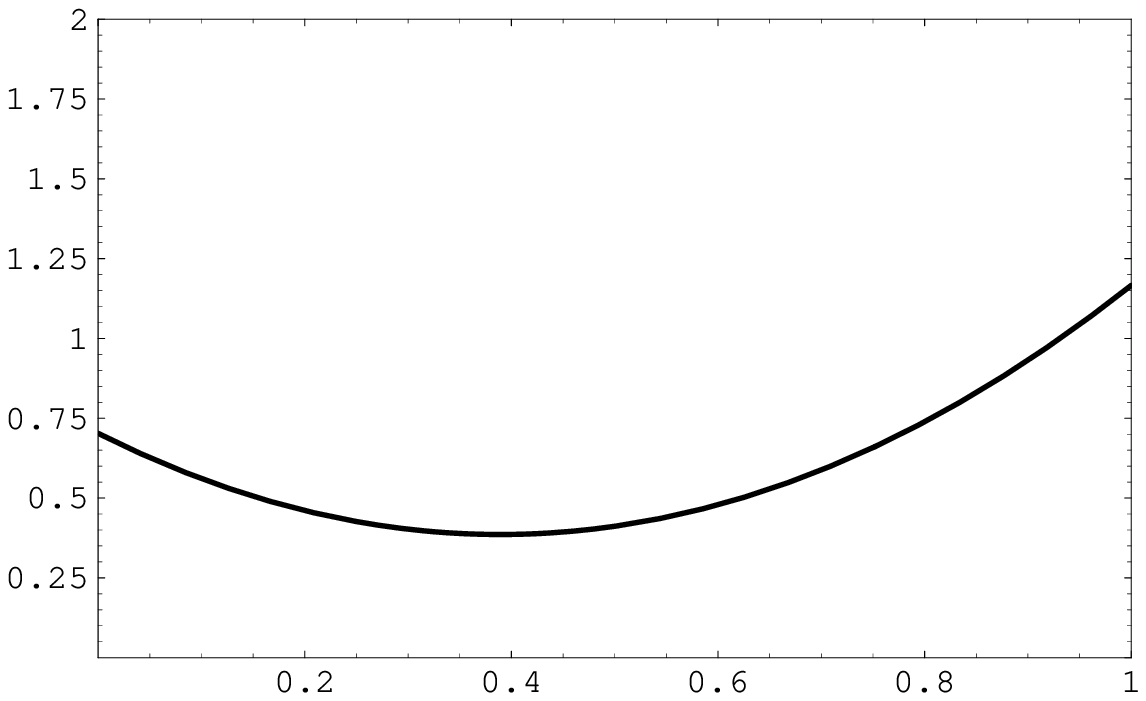}}\\
{\large $\sin^2\gamma$}\\ {\large\bfseries Fig.3b}
\end{center}


\newpage
\begin{thebibliography}{99}

\bibitem{1} N.Cabibbo, L.Maiani, G.Parisi, R.Petronzio, Nucl.Phys. {\bf B158} (1979) 295;
M.A.Beg, C.Panagiotakopolus, A.Sirlin, Phys.Rev.Lett. {\bf 52} (1984) 883;
M.Lindner, Z. Phys. {\bf C31} (1986) 295;
P.Q.Hung, G.Isidori, Phys.Lett. {\bf B402} (1997) 122;
T.Hambye, K.Reisselmann, Phys.Rev. {\bf D55} (1997) 7255.
\bibitem{211} M.Sher, Phys.Rep. {\bf 179} (1989) 273.
\bibitem{2} M.Lindner, M.Sher, H.W.Zaglauer, Phys.Lett. {\bf B228} (1989) 139;
N.V.Krasnikov, S.Pokorski, Phys.Lett. {\bf B288} (1991) 184;
M.Sher, Phys.Lett. {\bf B317} (1993) 159; ibid. {\bf B331} (1994) 448;
N.Ford, D.R.T.Jones, P.W.Stephenson, M.B.Einhorn, Nucl.Phys. {\bf B395} (1993) 17;
G.Altarelli, G.Isidori, Phys.Lett. {\bf B337} (1994) 14;
J.A.Casas, J.R.Espinosa, M.Quiros, Phys.Lett. {\bf B342} (1995) 171;
M.A.Diaz, T.A.Ter Veldius, T.J.Weiler, Phys.Rev. {\bf D54} (1996) 5855.
\bibitem{5} U.Amaldi, W.de Boer, H.F$\ddot u$rstenau, Phys.Lett. {\bf B260} (1991) 447;
P.Langaker, M.Luo, Phys.Rev. {\bf D44} (1991) 817;
J.Ellis, S.Kelley, D.V.Nanopoulos, Nucl.Phys. {\bf B373} (1992) 55.
\bibitem{6} A.G.Riess {\itshape et al.}, Astron.J. {\bf 116}, 1009 (1998);
S.Perlmutter {\itshape et al.}, Astrophys.J. {\bf 517}, 565 (1999).
\bibitem{7} N.Arkani--Hamed, S.Dimopoulos, G.Dvali, Phys.Lett. {\bf B429} (1998) 263;
I.Antoniadis, N.Arkani--Hamed, S.Dimopoulos, G.Dvali, Phys.Lett. {\bf B436} (1998) 257.
\bibitem{8} L.Randall, R.Sundrum, Phys.Rev.Lett. {\bf 83} (1999) 3370;
L.Randall, R.Sundrum, Phys.Rev.Lett. {\bf 83} (1999) 4690.
\bibitem{9} K.R.Dienes, E.Dudas, T.Cherghetta, Phys.Lett. {\bf B436} (1998) 55;
K.R.Dienes, E.Dudas, T.Cherghetta, Nucl.Phys. {\bf B537} (1999) 47.
\bibitem{10} N.Arkani--Hamed, S.Dimopoulos, N.Kaloper, R.Sundrum, Phys.Lett. {\bf B480} (2000) 193;
S.Kachru, M.Schulz, E.Silverstein, Phys.Rev. {\bf D62} (2000) 085003; G.Dvali, G.Gabadadze, M.Shifman,
Phys.Rev.{\bf D67} (2003) 044020.
\bibitem{11} D.L.Bennett, H.B.Nielsen, Int.J.Mod.Phys. A {\bf 9}, 5155 (1994);
{\it ibid} {\bf 14}, 3313 (1999);
D.L.Bennett, C.D.Froggatt, H.B.Nielsen, in {\itshape Proceedings of the 27th International Conference on
High energy Physics, Glasgow, Scotland, 1994}, p.557.
\bibitem{12} C.D.Froggatt, H.B.Nielsen, Phys.Lett. {\bf B368} (1996) 96.
\bibitem{13} C.D.~Froggatt and H.B.~Nielsen, Surv.~High Energy Phys. {\bf 18}, 55 (2003); C.D.~Froggatt, L.V.~Laperashvili and H.B.~Nielsen,
Int.~J.~Mod.~Phys. A {\bf 20}, 1268 (2005);
C.D.~Froggatt, arXiv:hep-ph/0412337.
\bibitem{14} S.W.Hawking, Phys.Rev. {\bf D37} (1988) 904; S.Coleman, Nucl.Phys. {\bf B307} (1988) 867;
{\it ibid} {\bf B310} (1988) 643; T.Banks, Nucl.Phys. {\bf B309} (1988) 493.
\bibitem{book}
C.D.Froggatt and H.B.Nielsen, {\it Origin of Symmetries}, (World Scientific, Singapore, 1991); D.L.~Bennett, C.D.~Froggatt and H.B.~Nielsen, 
{\it Perspectives in Particle Physics '94, World Scientific, 1995}, p.~255, ed. D.~Klabu\u{c}ar, I.~Picek and D.~Tadi\'{c} [arXiv:hep-ph/9504294]; 
C.D.Froggatt and H.B.Nielsen, arXiv:hep-ph/9607375.
\bibitem{15} C.D.Froggatt, L.V.Laperashvili, R.Nevzorov, H.B.Nielsen, Phys.Atom.Nucl. {\bf 67} (2004) 582.
\bibitem{151} C.D.Froggatt, R.Nevzorov, H.B.Nielsen, arXiv:hep-ph/0511259.
\bibitem{16} J.F.Gunion, H.E.Haber, G.Kane, S.Dawson, {\it The Higgs Hunter's Guide},
(Addison--Wesley, Redwood City, CA, 1990).
\bibitem{171} K.Inoue, A.Kakuto, Y.Nakano, Prog.Theor.Phys. {\bf 63} (1980) 234;
C.T.Hill, C.N.Leung, S.Rao, Nucl.Phys. {\bf B262} (1985) 517;
H.E.Haber, R.Hempfling, Phys.Rev. {\bf D48} (1993) 4280.
\bibitem{17} H.Komatsu, Prog.Theor.Phys. {\bf 67} (1982) 1177;
D.Kominis, R.S.Chivukula, Phys.Lett. {\bf B304} (1993) 152.
\bibitem{19} S.Glashow, S.Weinberg, Phys.Rev. {\bf D15} (1977) 1958.
\bibitem{24} M.Ciuchini, G. Degrassi, P. Gambino, G.F. Giudice, Nucl.Phys. {\bf B527} (1998) 21;
P.Gambino, M.Misiak, Nucl.Phys. {\bf B611} (2001) 338.
\bibitem{25} P.A.Kovalenko, R.B.Nevzorov, K.A.Ter--Martirosyan, Phys. Atom. Nucl.
{\bf 61}(1998) 812.
\bibitem{26} R.Flores, M.Sher, Ann.Phys. {\bf 148} (1983) 295;
A.Bovier, D.Wyler, Phys.Lett. {\bf B154} (1985) 43;
A.J.Davies, G.C.Joshi, Phys.Rev.Lett. {\bf 58} (1987) 1919;
J.Maalampi, J.Sirkka, I.Vilja, Phys.Lett. {\bf B265} (1991) 371;
S.Kanemura, T.Kubota, E.Takasugi, Phys.Lett. {\bf B313} (1993) 155;
A.G.Akeroyd, A.Arhrib, E.Naimi, Phys.Lett. {\bf B490} (2000) 119.
\bibitem{27} S.Nie, M.Sher, Phys.Lett. {\bf B449} (1999) 89.
\bibitem{29} C.D.Froggatt, H.B.Nielsen, Y.Takanishi, Phys.Rev. {\bf D64} (2001) 113014.
\bibitem{21} R.D.Peccei, H.R.Quinn, Phys.Rev.Lett {\bf 38} (1977) 1440;
Phys.Rev. {\bf D16} (1977) 1791.
\bibitem{28} S.Kanemura, T.Kasai, Y.Okada, Phys.Lett. {\bf B471} (1999) 182.

\end{thebibliography}
\end{document}